\documentclass[preprint, aps]{revtex4}

\usepackage{latexsym, graphicx}

\begin{document}

\title{Black Hole Information in a Detector (Atom) - Field Analog
  \footnote{Invited plenary talk at the workshop 
  ``From Quantum to Emergent Gravity: Theory and Phenomenology", 
  Trieste, Italy, June 11-15, 2007.}}
\author{B. L. Hu \footnote{Speaker.}}
\email{blhu@umd.edu}
\affiliation{Joint Quantum Institute and Maryland Center for
Fundamental Physics, Department of Physics, University of Maryland,
College Park, Maryland 20742-4111, USA}
\author{Shih-Yuin Lin}
\email{sylin@phys.cts.nthu.edu.tw}
\affiliation{Physics Division, National Center for Theoretical Science,
P.O. Box 2-131, Hsinchu 30013, Taiwan}
\begin{abstract}
This is a synopsis of our recent work \cite{LHRecoh} on quantum
entanglement, recoherence and information flow between an uniformly
accelerated detector and a massless quantum scalar field. The
availability of exact solutions to this model enables us to explore
the black hole information issue with some quantifiable results and
new insights.
To the extent this model can be used as an analog to the system of a
black hole interacting with a quantum field, our result seems to
suggest in the prevalent non-Markovian regime, assuming unitarity for
the combined system, that  black hole information is not lost but
transferred to the quantum field degrees of freedom. This combined
system will evolve into a highly entangled state between a remnant of
large area (in Bekenstein's black hole atom analog) without any
information of its initial state, while the quantum field is imbued
with complex information content not-so-easily retrievable by a local
observer.
\end{abstract}

\maketitle

\noindent {\bf A note on BLH's talk in this meeting}

This was meant to be a conference paper based on the invited talk of
BLH at the Workshop ``From Quantum to Emergent gravity: Theory and
phenomenology" 11-15 June, 2007 at SISSA, Trieste, Italy. The overall
theme of the talk is on a new view towards quantum gravity as a
theory of the microscopic structure of spacetime. The statement is
that such a theory may be inequivalent to that obtained by quantizing
general relativity. This highly successful theory for the macroscopic
structure of spacetime may just be an effective theory valid at the
low energy, long wavelength limit of the underlying theory describing 
the microscopic structures of spacetime. With the metric and connection
forms acting as the collective or hydrodynamic variables of the
microscopic theory, classical gravity in this view is emergent and
these variables will lose their meaning at shorter wavelengths and
higher energies. Examples are drawn from hydrodynamics, critical
dynamics, quantum fluids, and atomic - condensed matter physics to
illustrate how very different the emphasis and approaches, the goals
and methodology are between the traditional view of quantizing
general relativity and the new view of gravity as an emergent theory.
Instead of placing emphasis on quantization and posing the challenge
of finding the quantum version of a classical theory, the new
challenge is to infer the microscopic structure from the known
macroscopic phenomena. This has been the task for physicists for
centuries. For this, concepts and methods from nonequilibrium
statistical mechanics and examples from strongly correlated many-body
systems will probably play an essential role hitherto largely ignored
for quantum gravity.

From this general backdrop, special emphasis was placed in this
talk on the properties of emergent theories, distinguishing those
which can be logically and methodically deduced from a microscopic
theory and those which cannot, at least not without knowing in some
degree certain attributes of the macroscopic theory
\footnote{Hydrodynamics or nuclear physics are  examples of the
former (from molecular dynamics and QCD respectively) and quantum
Hall effect is often quoted as an example of the latter: One can
construct useful theories only AFTER such an effect was observed, but
not (or highly unlikely) BEFORE.}. The former type poses the
difficult task of inferring the unknown micro structures from the
known macro phenomena, but the latter type adds to it a more
difficult challenge of finding the unknown characteristics of an
emergent theory. It is not enough to say something is emergent
\footnote{Oftentimes referring to something as emergent is an
euphemism for saying we don't really know what the underlying
theories are. It borders on a `copt-out', one with a philosophical
tinge notwithstanding.} -- If we don't know what gives rise to the
emergent theory, at least we should try to describe the underlying
processes or mechanisms that could lead to such phenomena. These are
the new challenges of quantum gravity in the emergent vein.

The general theme of BLH's invited talk can be found ( $\sim 2/3$ of
the slides) in the website of the Loops '07 meeting, 25 - 30 June
2007 in Morelia, Mexico\footnote{{\tt
http://www.matmor.unam.mx/eventos/loops07/}}. The issues of nonlocality 
and stochasticity in the quantum-classical and micro-macro interfaces 
and how they bear on emergent gravity will be discussed in Ref. 
\cite{EmerGrav}.

Instead of repeating what can be found in published papers and essays
the speaker finds it more useful to present some of the newest
research results, as in this case, the work summarized in this paper.
This is deemed excusable or even appropriate,  because analog gravity
is an important theme in this Workshop.

\section{Detector-field System and Black-hole Atom Analog}

In this work we wish to understand the black hole information issue
\cite{Haw76b,Haw05,Page80} (see, e.g., \cite{Preskill93,Page94} for
an overview) by probing into one key aspect of it, namely, how
information is distributed between the black hole and the quantum
field throughout its history. We are not in a position to
unequivocally decide on the end state of a black hole after emitting
Hawking radiation \cite{Haw75} -- remnants, naked singularity, baby
universe formation or complete evaporation. (See e.g.,
\cite{Witten91,CGHS,RST,Wilczek93,PolStr}.) Nor are we equipped to
enter into the debate in whether there is net information lost in a
black hole and, the grander issue of whether unitarity in the laws of
physics is violated.

On the specific aspect we are interested in, i.e., where information
is registered, stored and transferred, a prevailing thought of the
school which upholds the validity of unitarity (and thus advocates no
information loss) is that information resides in the correlation
between the black hole and its Hawking radiation which persists down
to the very end of evaporation.   Here we want to examine with the
help of an analog model another contrasting view expressed earlier by
one of us \cite{HuErice} (see also \cite{Wilczek93}), assuming no
violation of unitarity, namely, that information in the black hole is
not lost but transferred and dispersed into the quantum field through
Hawking radiation, nor does it reside predominantly in the
correlations between the black hole and the quantum field. Ref.
\cite{HuErice} proposes to use the correlation functions of an
interacting field as registers of information and the dynamics of
correlations as a measure of information flow accompanied by a
suggested scenario where information in the black hole is transferred
to the quantum field. According to the viewpoint put forth there, the
appearance of information loss is primarily owing to the fact that
actual physical measurements by a local observer are limited in
accuracy, i.e., one can only access the lowest order correlation
functions, beginning with the mean field and the two-point functions.
It also highlights the huge capacity of a quantum field in storing
and dispensing information.

We have recently studied an exactly solvable model where an uniformly
accelerated detector is linearly coupled to a massless scalar field
initially in the Minkowski vacuum. Based on the results of this study
we drew some suggestive implications on the black hole information
flow issue by invoking Bekenstein's black hole atom analog.  We give
a brief description of the black hole atom analog below, followed by
the detector-field model we studied.

Bekenstein observed that the black hole behaves like an ensemble of
quantum mechanical atoms \cite{Bek75, BM76, Bek97}, whose spontaneous
emissions correspond to Hawking radiation, and the energy level of 
the atom is analogous to the area level of a black hole.
When black holes are fed with field quanta, they tend to absorb more 
than emit energy. Indeed, Bekenstein and Meisels showed that 
for black holes the Einstein B coefficient for stimulated absorption 
is greater than the coefficient for stimulated emission \cite{BM76}.
Following this idea the atom (or the particle detector considered 
below) itself can be treated as an analog of the black hole. 
We can learn some physics about the black hole information issue from
the ordinary quantum atom-field interacting system. (Note that the
theory for black hole atoms does not a priori assume any violation of
unitarity and hence no information loss.)

Quantum decoherence and recoherence of an inertial detector
interacting with a quantum field in the ultraweak coupling regime has
been studied before by Anglin {\it et al.} \cite{AngRecoh} They
claimed that, as soon as the coupling is switched on, the oscillator
loses the quantum coherence on a very short decoherence time
corresponding to the cut-off time scale. But almost all the quantum
coherence will recover in the end, after a much longer relaxation
time. They then used these findings to draw some implications on the
black hole information problem. Please refer to our recent paper
\cite{LHRecoh} for a comparison of our findings in this more general
case with theirs.

Let us consider a moving harmonic oscillator with internal degree of
freedom $Q$ (known as the Unruh-DeWitt(UD) detector \cite{Unr76,DeW79,BD})
interacting with a massless quantum scalar field $\Phi$ in
four-dimensional Minkowski space. The action of the combined
particle detector - quantum field system is given by \cite{LH2005}
\begin{equation}
  S=\int d\tau {m_0\over 2}\left[ \left(\partial_\tau Q\right)^2
    -\Omega_0^2 Q^2\right] -\int d^4 x {1\over 2}\partial_\mu\Phi
    \partial^\mu\Phi  +{\lambda_0}\int d\tau\int d^4 x Q(\tau)\Phi (x)
  \delta^4\left(x^{\mu}-z^{\mu}(\tau)\right), \label{Stot1}
\end{equation}
where $\lambda_0$ is the coupling constant, $m_0$ and $\Omega_0$ are
the bare mass and natural frequency of the detector, respectively. We
will consider the cases when it is uniformly accelerated along the
trajectory $z^\mu(\tau)=(a^{-1}\sinh a\tau, a^{-1}\cosh a\tau,0,0)$
with proper acceleration $a$. For the cases of detectors at rest
($a=0$), we have learnt in \cite{LH2005, LH2006} that the two-point
functions of our UD detector theory in (3+1)D with finite $a$ have no
singular behavior as $a\to 0$. Hence all our results expressed in
terms of these two-point functions apply equally well to the case of
detectors at rest.

\section{Information flow between the detector and the quantum field}
\label{EvsC}

We study the case when the initial state of the combined system is a
direct product of a quantum state $\left|\right. q\left.\right>$ for
the detector $Q$ and the Minkowski vacuum $\left|\right. 0_M
\left.\right>$ for the field $\Phi$,
\begin{equation}
  \left|\right. \psi(\tau_0)\left.\right> =
  \left|\right. q\left.\right> \otimes
  \left|\right. 0_M \left.\right>. \label{initstat}
\end{equation}
Since the combined system is linear, the operators evolve in the
Heisenberg picture as linear transformations.
When sandwiched by the factorized initial state $(\ref{initstat})$,
the two-point functions of the detector and those of the field split
into two parts \cite{LH2005}, e.g.,
\begin{equation}
  \left<\right. Q(\tau)Q(\tau')\left.\right>  =
  \left<\right.q \,|\, q\left.\right>\left< \right. Q(\tau)Q(\tau')
  \left.\right>_{\rm v}+ \left< \right.Q(\tau)Q(\tau')\left.
  \right>_{\rm a}\left< 0_M| 0_M\right>. \label{splitQQ}
\end{equation}
Here $\left<\right. ..\left.\right>_{\rm a}$ depends on the initial
state of the detector only, while $\left<\right. ..\left.
\right>_{\rm v}$ depends on the initial state of the field, namely
the Minkowski vacuum. Therefore by studying the correlation functions
$\left<\right. .. \left.\right>_{\rm a}$ of the detector and those of
the field, one can monitor how the information initially in the
detector is flowing into the field.

Indeed, from Ref. \cite{LH2005}, we learned that $\left<\right.
Q^2\left. \right>_{\rm a}$, $\left<\right.P^2\left.\right>_{\rm a}$
and $\left<\right.P,Q\left.\right>_{\rm a}$ all decay after the
coupling is switched on, and the information about the initial state
of the detector is subsumed into the quantum field (in
$\left<\right.\Phi(x)\Phi(x')\left. \right>_{\rm a}$, etc.). This
view is further supported by the energy conservation law found in
\cite{LH2005} between the internal energy of the detector and the
radiated energy of a monopole. The energy in
$\left<\right.Q^2\left.\right>_{\rm a}$ and
$\left<\right.P^2\left.\right>_{\rm a}$ will be converted to monopole
radiation while the state of the detector at late times is sustained
only by the vacuum fluctuations of the field.

\section{Entanglement between the detector and the field}

For a bipartite system with the combined system in a pure state,
such as the Unruh-DeWitt detector theory with initial state
$(\ref{initstat})$, the purity of each sub-system
\begin{equation}
 {\cal P}\equiv Tr_Q \left[\rho^R(Q,Q')\right]^2 =
 {\cal P}_\Phi \equiv Tr_\Phi\left(\rho^R [\Phi,\Phi']\right)^2
\end{equation}
is equal to the other and gives a measures of the entanglement
between them. On the other hand, the purity of a two-level atom is
proportional to its polarization, thus providing a measure of quantum
coherence in that atom. Here we extend this view to our system and
use the value of the purity function as a measure of quantum
coherence in the detector and in the field as well.

The behavior of quantum coherence ``flow" is quite different from
energy flow. When the coupling is switched on, both the quantum
coherence in the detector and the quantum coherence in the field
decrease, while the entanglement between them increases. So quantum
coherence does not flow from one subsystem to the other; It goes into
sustaining the entanglement between the two subsystems.

A lower purity means lesser quantum coherence in the detector or the
field and stronger entanglement between the detector and the field.
To show this one may define the linear entropy in terms of the purity
as
\begin{equation}
  S_L \equiv 1-{\cal P} . \label{LiEtp}
\end{equation}
Now the value of $S_L$ is zero for a detector in a pure state and
is positive for a detector in a mixed state. By definition the linear
entropy seen by the detector will be equal to the linear entropy
seen by the field. Also, the greater the von Neumann entropy of the
detector, the greater $S_L$. Thus $S_L$ could serve as a measure of
entanglement between the detector and the field, just as good as the
von Neumann entropy, for the bipartite system in a pure state.

As an example, suppose the detector is in a cat state at the
initial moment $\tau_0$,
\begin{equation}
  \left|\right. q(\tau_0)\left.\right> =
  \cos \varphi\left|\right. E_0\left.\right>+
  e^{i\delta}\sin\varphi\left|\right. E_1\left.\right>,
  \label{initcat}
\end{equation}
where $\left|\right. E_0\left. \right>$ and $\left|\right.
E_1\left.\right>$ are the ground state and the first excited state of
the free detector, $\varphi$ is the mixing angle and $\delta$ is a
constant phase. The reduced density matrix (RDM) of the detector for this
initial state reads
\begin{eqnarray}
  \rho^R(Q,Q';\tau) &=& \sqrt{G^{11}+G^{22}+2G^{12}\over\pi}
    e^{-G^{ij}Q_i Q_j} \left\{\cos^2\varphi +
    \sin^2\varphi(C +A^{ij}Q_i Q_j ) + \right.\nonumber\\ & & \left.
  \sin\varphi\cos\varphi\left[(e^{i\delta}B^1+e^{-i\delta}B^{2*})Q+
  (e^{-i\delta}B^{1*}+e^{i\delta}B^2) Q'\right]
  \right\}, \label{RDMD}
\end{eqnarray}
where $i,j =1,2$, $Q_i = (Q, Q')$. The coefficients $C$, $A^{ij}$, $B^j$,
$G^{ij}$ and $F$ could be expressed in terms of the two-point correlation
functions of the detector. Explicit expressions of the two-point functions
needed here have been listed in Appendix A of Ref. \cite{LH2006}.
Actually all $A^{ij}$ and $B^j$ will vanish at late times
($\gamma\eta\gg 1$  with $\gamma\equiv \lambda_0^2/8\pi m_0$ and
$\eta\equiv \tau-\tau_0$) when the RDM of all choices of $\delta$ and
$\varphi$ for initial states become a universal one,
\begin{equation}
  \left.\rho^R\right|_{\gamma\eta \gg 1}= \left.\rho^R
  \right|_{\varphi=0, \gamma\eta \gg 1}. \label{lateRhoR}
\end{equation}
Hence the purity (or the linear entropy) goes to a universal value at
late times.

\begin{figure}
\includegraphics[width=7.3cm]{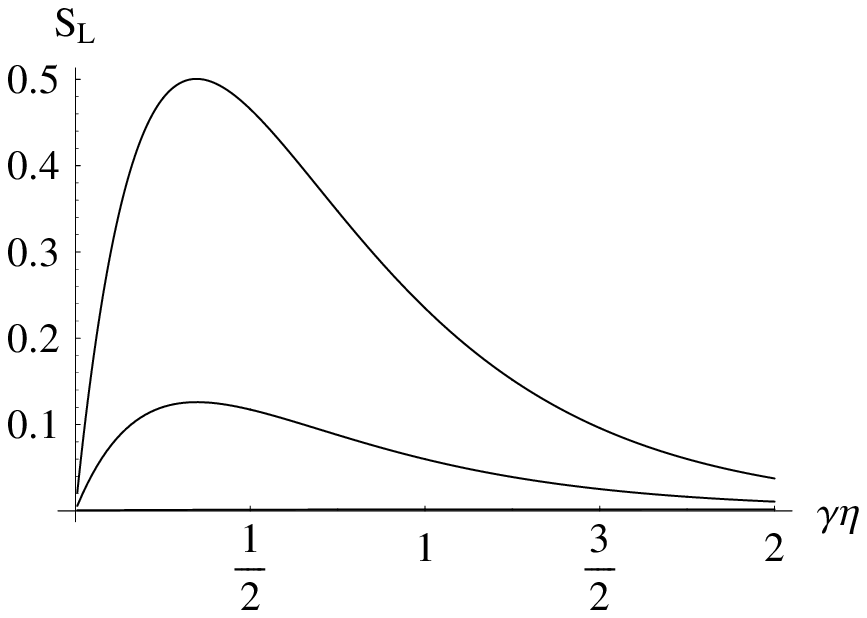}
\includegraphics[width=7.3cm]{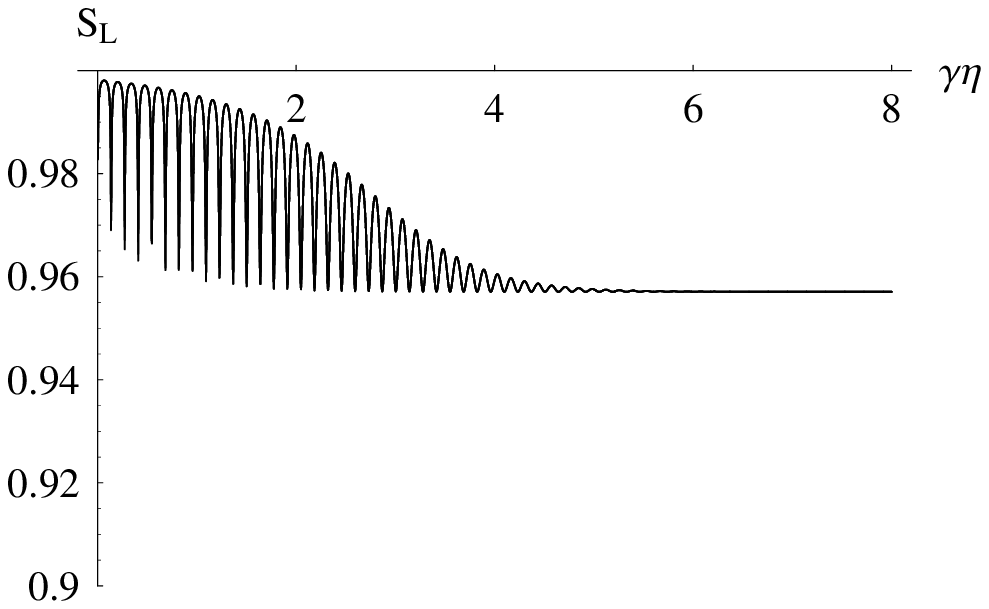}
\caption{Evolution of linear entropies $S_L$ of $(3.4)$ in detector's
proper time. Here $\Omega =2.3$, $a=2$, $m_0=\hbar=1$,
$\Lambda_1=\Lambda_0=10000$, $\delta=0$, $\gamma = 10^{-7}$(left) and
$0.1$(right). The three curves from top to bottom in each plot (the
bottom one in the left plot is very close to the $\gamma\eta$ axis,
while the three curves are indistinguishable in the right plot) have
$\varphi=\pi/2, \pi/4, 0$, respectively. } 
\label{SL2}
\end{figure}

\subsection{recoherence in the ultraweak coupling regime}

The linear entropies of the RDM $(\ref{RDMD})$ with different
parameters are illustrated in Figure \ref{SL2}. In the ultraweak
coupling regime ($\gamma \Lambda_1 \ll a,\Omega$, where $\Lambda_1$
is a large constant corresponding to the time-resolution or the
frequency cut-off of this theory, $\Omega\equiv \sqrt{\Omega_r^2
-\gamma^2}$ with  $\Omega_r$ the renormalized natural frequency of
the detector) \cite{LH2006}, for $a/\Omega$ sufficiently small, one
has
\begin{equation}
   {\cal P} \approx 1+2
   e^{-2\gamma\eta}\left(e^{-2\gamma\eta}-1\right)\sin^4\varphi .
\label{SE1to0}
\end{equation}
One can see that in this regime ${\cal P}$
is very close to unity at late times, when each subsystem re-gains
almost all quantum coherence and turns into a nearly pure state.

Indeed, observing the left plot of Figure \ref{SL2}, the linear
entropy $S_L$ in the detector
increases from zero right after the coupling is switched on, reaches
a maximum $(1/2)\sin^4\varphi$ at $\eta \approx \ln 2/2\gamma$, then
decays to a small common value that detectors with all other initial
states will asymptopte to. This decay of the degree of entanglement
or the restoration of the degree of quantum coherence is known as
``recoherence" \cite{AngRecoh}. 
The late-time recoherence manifests only in the ultraweak coupling
regime with sufficiently low acceleration (temperature), where the
late-time RDM of the detector looks very close to the density matrix
of the ground state of a free detector. Thus the recoherence here
characterizes the process of spontaneous emission by which the
detector initially in an excited state will finally fall into a
steady state which is very close to the ground state of the free
detector. Nevertheless, {\it full} recoherence is impossible once the
coupling is on, since the late-time linear entropy
\begin{equation}
  S_L|_{\gamma\eta\to \infty} \approx 1-\tanh{\pi\Omega\over a} +
  \gamma {2\tanh^2{\pi\Omega\over a}\over \pi \Omega}\left[
  \Lambda_1 + \ln{\Omega\over a} -  {\rm Re}\left[ \psi\left(
  {i\Omega\over a}\right) + {i\Omega\over a}\psi^{(1)}\left(
  {i\Omega\over a}\right)\right]\right] + O(\gamma^2)
\end{equation}
($\psi^{(1)}(x)\equiv d\psi(x)/dx$) remains nonzero for any positive
$\gamma$, even when $a\to 0$.

If the detector is initially in its first excited state
($\varphi=\pi/2$), the two-point function $\left<\right. Q^2 \left.
\right>$ at $\eta=\ln 2/2\gamma$ will have the same value as the
average of those $\left<\right. Q^2(\tau_0) \left.\right>$ from the
ground state and from the first excited state at the initial moment.
It may seem that the intermediate state of the detector during the
transition is a cat state which is a superposition of the ground
state and the first excited state of the detector, but this is not
true.
Large $S_L$ in transient indicates that the intermediate state during
the spontaneous emission is a mixed state, in contrast to the zero
$S_L$ for a pure cat state (at the initial moment of the middle curve
in the left plot of Figure \ref{SL2}). The value of
$\left<\right. Q^2\left.\right>$ in transient is mainly an
ensemble (probabilistic) average of the population in the ground
state and the population in the first excited state.

\subsection{Beyond the ultraweak coupling regime}
\label{nMark}

When $\Omega \Lambda_1 \gg a, \gamma$, the system is in the
non-Markovian regime and the purity is always small (right plot in
Figure \ref{SL2}), which implies that the detector experiences strong
decoherence associated with strong entanglement between the detector
and the field. The behavior of the detector is dominated by the
physical cut-offs and the differences between various initial states
of the detector can be negligible. For example, when $\varphi=\pi/2$,
just like the case with the detector initially in its ground state,
the initial distribution of the RDM in the energy-eigenstate
representation $\rho^R_{m,n}$ peaked at the element $\rho^R_{1,1}$
would, upon the switch-on of the coupling, collapse rapidly into a
distribution widely spread over the whole density matrix, for which
the energy eigenstates of the free detector cannot form a good basis
because the off-diagonal terms of the RDM do not vanish even in a
steady state \cite{LH2006}. The late-time linear entropy
\begin{equation}
  S_L|_{\gamma\eta\gg 1} \approx 1- {\pi/2  \over
  \sqrt{ {2\over\Omega}\gamma\Lambda_1 {\rm Re}
  \left[ {ia \over \gamma+ i\Omega} -2i\psi_{\gamma+i\Omega}
  \right]}} + O(\Lambda_1^{-3/2})
\end{equation}
is very close to unity. Hence there is no late-time recoherence in
this regime, where the quantum state of the combined system is far
from being a direct product of the state of the detector and that of
the field.

\begin{figure}
\includegraphics[width=7.3cm]{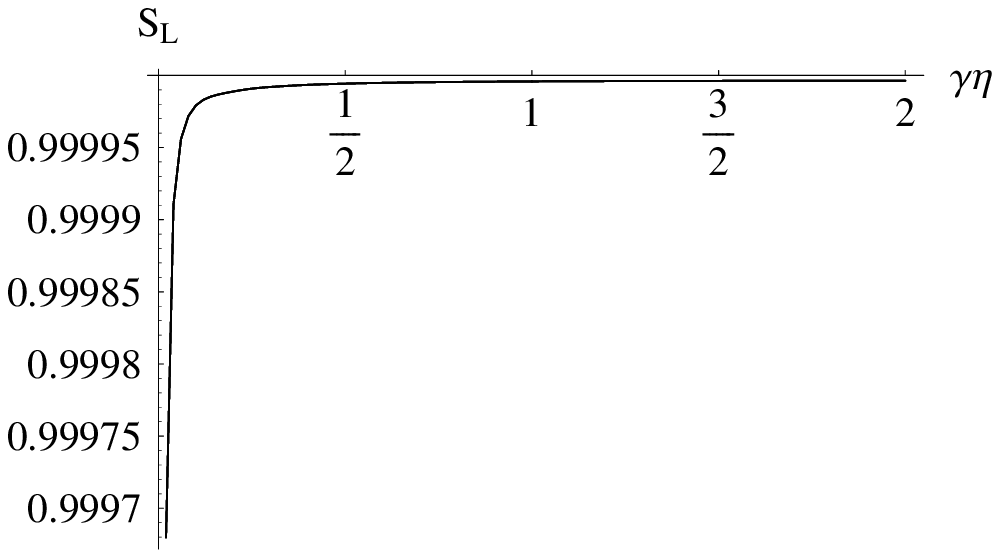}
\includegraphics[width=7.3cm]{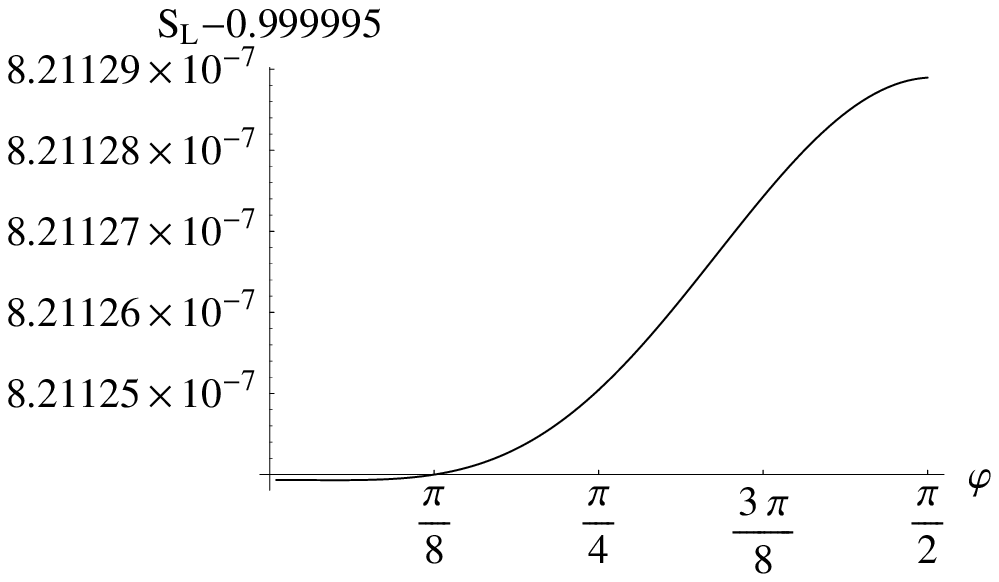}
\caption{$S_L$ in ultrahigh acceleration limit: $a=2\times 10^6$,
  $\gamma =0.1$, other parameters have the same values as those in
  Figure 1. As indicated in the right plot, the curves with
  different $\varphi$ are not distinguishable in the left plot.}
\label{higha}
\end{figure}

A large linear entropy at late times also shows up at the ultrahigh
acceleration (or Unruh temperature) limit ($a\gg\gamma\Lambda_1$,
$\Omega$),
\begin{equation}
 S_L|_{\gamma\eta\gg 1}\approx 1-{\pi \Omega_r\over 2a} + O(a^{-2}),
\end{equation}
(see Figure \ref{higha}). When the coupling is weak enough, the energy
eigenstates can still form a good basis and the RDM of the detector
is approximately a thermal state in the energy-eigenstate
representation: all off-diagonal elements are negligible. Note that
the ultrahigh temperature limit is still in a Markovian regime, so we
see that strong entanglement does not imply a non-Markovian process.

\section{Discussion}

From our results we can see that, as long as the coupling between
the detector and the field is on, the detector and the field are
separately in a mixed state of its own, while the combined system
remains in a pure state. The mixed state of the detector carries
information of the initial state until $\left< Q^2\right>_a$, $\left<
P^2\right>_a$ and $\left< P,Q\right>_a$ all decay away and the
detector reaches a steady state. At late times, while the energy
eigenstates of the free detector cannot form a good basis, the RDM of
the detector can be diagonalized to a Boltzmann distribution from
which one can read off the same effective temperature as those
reported in \cite{LH2006}. In this sense the final mixed state of the
detector is a thermal state containing no initial information. This
is true for both inertial and uniformly accelerating detectors, the
former case might be a surprise.

Similarly, when the black hole is radiating, the black hole itself
and the field outside the black hole are each in a mixed state. Only
in the ultraweak coupling limit can they restore most of their purity
at late times. Otherwise, in the more prevalent non-Markovian regime,
the area eigenstates of the black hole cannot form a good basis, and
the entanglement between the black hole and the field is always
large. Nevertheless, in this scenerio the quantum state of the
combined system is always pure due to the unitarity we assumed.

The existence of Einstein A and B coefficient for black holes
\cite{BM76} suggests that, if the field is initially in a vacuum
state, the information about the black hole would be encoded in its
spontaneous emission, namely, its emitted radiation which is not
exactly thermal \footnote{Note that as different from atoms, for a
real black hole its event horizon would have an effect on such
emission.}. All initial information in the black hole will eventually
go to the field at late times, while the final state of the black
hole is sustained by the vacuum fluctuations of the quantum field.
This is consistent with the ``no-hiding theorem" of Braunstein and
Pati (with their ancilla as our quantum field) \cite{BraPat}: no
information is hidden in the correlations between the field and the
black hole.

In our model the difference between the degrees of freedom of the
detector and those of the field  is put in by hand and never
disappears, so we cannot address whether the ground state corresponds
to a black hole remnant, an ordinary localized  mass, or nothing.
However, in Bekenstein's atom analog highly excited eigenstates
correspond to large-area black holes. Our results indicate that in
the rather general non-Markovian regime, the combined system would
evolve to a highly entangled state between the black hole and the
field, and the final state of the black hole would be a 
mixed state distributed widely from the ground state to the highly
excited area-eigenstates. So at late times in the broad ranged
non-Markovian regime the black hole could end up as a large remnant
with all its initial information already leaked out and
dispersed into the quantum field.\\

\noindent{\bf Acknowledgement}  BLH thanks the organizers of this
workshop, especially Stefano Liberati, for their invitation and
hospitality. Part of the work reported here was done when he visited
the Institute for Advanced Study, Princeton in Spring '07.
SYL wishes to thank Zheng-Yao Su for pointing out the conservation
relation between quantum coherence and entanglement.
This work is supported in part by an NSF Grant PHY-0601550.


\begin{thebibliography}{99}

\bibitem{LHRecoh}
S.-Y. Lin and B. L. Hu, {\it Quantum entanglement,  recoherence and
information flow in a particle-field system: implications for black
hole information issue} [{\tt arXiv:0710.0435}].

\bibitem{EmerGrav} B. L. Hu, {\it Nonlocality and Stochasticity in 
Emergent Gravity: Issues at the Micro-Macro and Quantum-Classical 
Interfaces}, Talk at {\it Peyresq 12}, \emph{Class. Quant. Gravity.} 
(Special Issue).

\bibitem{Haw76b} S. W. Hawking, {\it Breakdown of predictablity in
gravitational collapse}, \emph{Phys. Rev. D} {\bf 14}, 2460 (1976).

\bibitem{Haw05} S. W. Hawking, {\it Information loss in black holes},
\emph{Phys. Rev. D} {\bf 72}, 084013 (2005).

\bibitem{Page80} D. N. Page, {\it Is information lost down black holes?},
in {\it Proceedings of the 9th International Conference on General
Relativity and Gravitation}, edited by E. Schmutzer
(Friedrich Schiller University, Jena, 1980).

\bibitem{Preskill93} J. Preskill, {\it Do black holes destroy
information?} in {\it Black Holes, Membranes, Wormholes and
Superstrings}, edited by K. Kalara and D. V. Nanopoulos
(World Scientific, Singapore, 1993) [{\tt hep-th/9209058}].

\bibitem{Page94} D. N. Page, {\it Black hole information}, in
{\it Proceedings of 5th Canadian Conference on General Relativity and
Relativistic Astrophysics}, edited by R. B. Mann and R. G. McLenaghan
(World Scientific, Singapore, 1994) [{\tt hep-th/9305040}].

\bibitem {Haw75} S. W. Hawking, {\it Particle creation by black holes},
\emph{Commun. Math. Phys.} {\bf 43}, 199 (1975).

\bibitem{Witten91} E. Witten, {\it String theory and black holes},
\emph{Phys. Rev. D} {\bf 44}, 314 (1991).

\bibitem {CGHS} C. G. Callan, S. B. Giddings, J. A. Harvey, and
A. Strominger, {\it Evanescent black holes}, \emph{Phys. Rev. D}
{\bf 45}, R1005 (1992) [{\tt hep-th/9111056}].

\bibitem{RST} J. G. Russo, L. Susskind, and L. Thorlacius,
{\it Black hole evaporation in (1+1)-dimensions}, \emph{Phys. Lett.}
{\bf B292}, 13 (1992)  [{\tt hep-th/9201074}];
{\it End point of Hawking radiation},
\emph{Phys. Rev. D} {\bf 46}, 3444 (1992) [{\tt hep-th/9206070}].

\bibitem{Wilczek93}
C. Holzhey and F. Wilczek, {\it Black Holes as Elementary Particles},
\emph{Nucl. Phys.} {\bf B380}, 447 (1992) [{\tt hep-th/9202014}];
F. Wilczek, {\it Quantum Purity at a Small Price: Easing a Black Hole
Paradox}, in {\it Black Holes, Membranes, Wormholes and Superstrings},
edited by K. Kalara and D. V. Nanopoulos (World Scientific, Singapore,
1993) [{\tt hep-th/9302096}].

\bibitem{PolStr} J. Polchinski and A. Strominger, {\it Unitary Rules
for Black Hole Evaporation}, \emph{Phys. Rev. D} {\bf 50}, 7403 (1994)
[{\tt hep-th/9410187}].

\bibitem{HuErice} B. L. Hu, {\it Correlation Dynamics of Quantum
Fields and Black Hole Information Paradox}, Erice Lectures, Sept. 1995,
in {\it String Gravity and Physics at the Planck Energy Scale},
edited by N. Sanchez and A. Zichichi (Kluwer, Dordrecht, 1996)
[{\tt gr-qc/9511075}].

\bibitem{Bek75} J. D. Bekenstein, {\it Statistical black-hole
thermodynamics}, \emph{Phys. Rev. D} {\bf 12}, 3077 (1975).

\bibitem{BM76}
J. D. Bekenstein and A. Meisels, {\it Einstein A and B coefficients
for a black hole}, \emph{Phys. Rev. D} {\bf 15},2775 (1976).

\bibitem{Bek97}
J. D. Bekenstein, {\it Quantum black holes as atoms}, in
{\it Prodeedings of the Eight Marcel Grossmann Meeting}, edited by
T. Piran and R. Ruffini (World Scientific, Singapore, 1999)
[{\tt gr-qc/9710076}].

\bibitem{AngRecoh} J. R. Anglin, R. Laflamme, W. H. Zurek and
J. P. Paz, {\it Decoherence and recoherence in an analogue of the
black hole information paradox}, \emph{Phys. Rev. D} {\bf 52},
2221 (1995) [{\tt gr-qc/9411073}].

\bibitem{Unr76} W. G. Unruh, {\it Notes on black-hole evaporation},
\emph{Phys. Rev. D} {\bf 14}, 870 (1976).

\bibitem{DeW79} B. S. DeWitt, {\it Quantum gravity: The new synthesis},
in {\it General Relativity: an Einstein Centenary Survey}, edited by
S. W. Hawking and W. Israel (Cambridge University Press, Cambridge,
1979).

\bibitem{BD} N. D. Birrell and P. C. W. Davies, {\it Quantum Fields in
Curved Space} (Cambridge University Press, Cambridge, 1982).

\bibitem{LH2005}
S.-Y. Lin and B. L. Hu, {\it Accelerated detector - quantum field
correlations: From vacuum fluctuations to radiation flux},
\emph{Phys. Rev. D} {\bf 73}, 124018 (2006) [{\tt gr-qc/0507054}].

\bibitem{LH2006}
S.-Y. Lin and B. L. Hu, {\it Backreaction and Unruh effect: New
insights from exact solutions of uniformly accelerated detectors},
\emph{Phys. Rev. D} {\bf 76}, 064008 (2007) [{\tt gr-qc/0611062}].

\bibitem{BraPat} S. L. Braunstein and A. K. Pati, {\it Quantum
information cannot be completely hidden in correlations:
Implications for the black-hole information paradox},
\emph{Phys. Rev. Lett.} {\bf 98}, 080502 (2007).

\end{thebibliography}
\end{document}